\documentclass[12pt, centerh1]{article}
\usepackage{setspace}

\usepackage{graphicx,colonequals}
\usepackage{natbib}
\usepackage{url}
\usepackage{amsmath}
\usepackage{amsfonts}
\usepackage{amssymb,amsthm}
\usepackage{url}
\usepackage{color}
\usepackage{amssymb,multirow,booktabs,graphicx}
\usepackage{graphics,psfrag,amsmath,amsthm,xr}
\usepackage{colonequals}
\usepackage{color}

\newcommand{\bx}{\mathbf{x}}

\newcommand{\bX}{\mathbf{X}}

\newcommand{\bV}{\mathbf{V}}

\newcommand{\cL}{\mathcal{L}}

\newcommand{\EE}{\mathbb{E}}

\newcommand{\bmu}{\mbox{\boldmath $\mu$}}

\newcommand{\bpsi}{\mbox{\boldmath $\psi$}}
\newcommand{\boeta}{\mbox{\boldmath $\eta$}}

\newcommand{\balpha}{\mbox{\boldmath $\alpha$}}

\newcommand{\btheta}{\mbox{\boldmath $\theta$}}
\newcommand{\bvartheta}{\mbox{\boldmath $\vartheta$}}

\newcommand{\bSigma}{\mbox{\boldmath $\Sigma$}}
\newcommand{\bGamma}{\mbox{\boldmath $\Gamma$}}

\newcommand{\bTheta}{\mbox{\boldmath $\Theta$}}
\newcommand{\bDelta}{\mbox{\boldmath $\Delta$}}

\newcommand*{\edit}{\textcolor{black}}

\title{Model-based clustering and classification using mixtures of multivariate skewed power exponential distributions}
\author{Utkarsh J.\ Dang$^*$ \qquad Michael P.\ B.\ Gallaugher$^{**}$\\ Ryan P.\ Browne$^{\dagger}$ \qquad\qquad Paul D. McNicholas$^{\dagger\dagger}$}
\date{\small $^*$Department of Health Sciences,
Carleton University, Ontario, Canada.\\ 
$^{**}$Department of Statistical Science, Baylor University, Texas, USA.\\
$^{\dagger}$Department of Statistics \& Actuarial Sciences, University of Waterloo, Ontario, Canada.\\
$^{\dagger\dagger}$Department of Mathematics \& Statistics, McMaster University, Ontario, Canada.}

\begin{document}

\maketitle

\begin{abstract}
Families of mixtures of multivariate power exponential (MPE) distributions have already been introduced and shown to be competitive for cluster analysis in comparison to other mixtures of elliptical distributions, including mixtures of Gaussian distributions. A family of mixtures of multivariate skewed power exponential distributions is proposed that combines the flexibility of the MPE distribution with the ability to model skewness. These mixtures are more robust to variations from normality and can account for skewness, varying tail weight, and peakedness of data. A generalized expectation-maximization approach, which combines minorization-maximization and optimization based on accelerated line search algorithms on the Stiefel manifold, is used for parameter estimation. These mixtures are implemented both in the unsupervised and semi-supervised classification frameworks. Both simulated and real data are used for illustration and comparison to other mixture families. 
\end{abstract}


\section{Introduction}
\label{sec:introduction}
Mixture modeling has been firmly established in the literature as a useful method for finding homogeneous groups within heterogeneous data. Using mixture models for cluster analysis has a long history \citep{hasselblad1966,day1969} dating at least to \citet{wolfe65}, who used a Gaussian mixture model for clustering. When using mixture models for clustering, which is known as model-based clustering, mixture models are used to partition data points to learn group memberships, or labels, of observations with unknown labels. If some observations are a~priori labeled, a semi-supervised analogue of model-based clustering is used and this is known as model-based classification. Extensive details on model-based clustering and classification are given by \cite{mcnicholas16a} and recent reviews are provided by \cite{bouveyron14} and \cite{mcnicholas16b}.

A $G$-component finite mixture model assumes that a random vector $\bX$ has density of the form
$$
f(\bx|\bvartheta)=\sum_{g=1}^G\pi_gf_g(\bx|\btheta_g),
$$
where $g=1,\ldots, G$, $\pi_g>0$ are the mixing proportions with $\sum_{g=1}^G\pi_g=1$, and $f_g(\cdot)$ are the component densities. 
The Gaussian mixture model \citep[see, e.g.,][]{banfield1993, celeux1995,tipping99b,mcnicholas08} remains popular due to its mathematical tractability. However, it is inflexible in the presence of cluster skewness and different levels of cluster kurtosis, and has been known to result in an overestimate of the number of clusters and poor density estimation for known clusters \citep[see][for examples]{franczak2014,dang15}. Therefore, it has become popular to consider mixtures of more flexible distributions for clustering to deal with such scenarios. 

Mixture models that can deal with varying cluster tail-weight, skewness and/or concentration, and kurtosis are increasingly becoming common. A small selection of such models include \edit{mixtures using power transformations \citep{zhu2022mattransmix}}, mixtures of multivariate $t$-distributions \citep{peel00,andrews2012}, mixtures of normal inverse Gaussian distributions \citep{karlis09,subedi2014,ohagan16}, mixtures of skew-$t$ distributions \citep{lin10,murray14b, vrbik2014, lee2014, lee2016},  mixtures of shifted asymmetric Laplace distributions \citep{morris13b,franczak2014}, mixtures of multivariate power exponential distributions \citep{dang15}, mixtures of variance-gamma distributions \citep{smcnicholas17}, and mixtures of generalized hyperbolic distributions and variations thereof \citep{browne15,murray17}. 

Two common approaches to introducing skewness are by means of a normal variance-mean mixture model, and via hidden truncation using an elliptical distribution and a skewing function. The former assumes that a random vector $\bX$ can be written in the form
$$\bX=\bmu+W\balpha+\sqrt{W}\bV,$$
where $\bmu$ and $\balpha$ are location and skewness vectors, respectively, $\bV\sim \mathcal{N}({\bf 0},\bSigma)$, $W\perp\bV$, and $W>0$ is a positive random variable with density $h(w|\bTheta)$. Depending on the distribution of $W$, different skewed distributions can be derived, e.g., the generalized hyperbolic, skew-$t$, variance-gamma and normal inverse Gaussian distributions. The hidden truncation approach makes use of a combination of an elliptical distribution and a skewing function. For example, a random vector $\bX$ follows a multivariate skew-normal distribution with skewness $\balpha$ if its density can be written as
$$
f(\bx)=2\phi_p(\bx|\bmu,\bSigma)\Phi(\balpha'\bx),
$$
where $\phi_p(\cdot)$ is the density of the p-dimensional normal distribution and $\Phi(\cdot)$ denotes the cumulative distribution function of the standard normal distribution \citep{azzalini96}.

The multivariate power exponential (MPE) distribution \citep{gomez1998} has been used in many different applications \citep[e.g.,][]{lindsey1999, cho2005, verdoolaege2008} and was recently used in the mixture model context by \citet{dang15}. Depending on the shape parameter $\beta$, either a leptokurtic or platykurtic distribution can be obtained. Specifically, if $\beta\in(0,1)$ then the distribution is leptokurtic, which is characterized by a thinner peak and heavy tails compared to the Gaussian distribution. If $\beta>1$, a platykurtic distribution is obtained, which is characterized by a flatter peak and thin tails compared to the Gaussian distribution. Other distributions can also be obtained for specific values of the shape parameter, for example, for $\beta=0.5$,  the distribution is a Laplace (double-exponential) distribution and, for $\beta=1$, it is a Gaussian distribution. Furthermore, when $\beta \rightarrow \infty$, the MPE becomes a multivariate uniform distribution. 

\cite{dang15} derived a family of mixtures of MPE distributions but those mixtures could only account for elliptical clusters. Previously, skew power exponential distributions have been discussed in the univariate case with constrained $\beta$ \citep{azzalini1986, diciccio2004,FERREIRA2011} or in the multivariate case as scale mixture of skew-normal with constrained $\beta$ \citep{branco2001}. Herein, we present mixtures based on a novel multivariate skewed power exponential (MSPE) distribution. As compared to earlier proposals, this distribution is more suitable for clustering and classification purposes and can be used for a wide range of $\beta$ (heavy, Gaussian, and light tails). Using an eigen-decomposition of the component scale distributions \citep[\`{a} la][]{celeux1995}, we construct a family of 16 MSPE mixture models for use in both clustering and semi-supervised classification. These models can account for varying tail weight (heavy, Gaussian, or light), peakedness (thinner or thicker than Gaussian), and skewness of mixture components.

\section{Background}
\label{sec:back}

Using the parametrization given by \cite{gomez1998}, a random vector $\bX$ follows a $p$-dimensional power exponential distribution if the density is
\begin{equation} \label{mpedef}
f(\bx|\bmu,\bSigma,\beta)=\frac{p\Gamma\left(\frac{p}{2}\right)}{\pi^{p/2}\Gamma\left(1+\frac{p}{2\beta}\right)2^{1+\frac{p}{2\beta}}}|\bSigma|^{-\frac{1}{2}} \exp\left\{ -\frac{1}{2} \delta(\bx)^\beta \right\},
\end{equation}
where $\bmu$ is the location parameter, $\bSigma$ is a scale matrix, $\beta$ determines the kurtosis, and $$\delta(\bx)\colonequals\delta \left(\bx|\bmu,\bSigma \right)=\left (\bx-\bmu \right)' \bSigma^{-1} \left(\bx-\bmu \right).$$ 
In a similar manner to \cite{azzalini96}, \cite{lin14} derived the multivariate skew $t$-normal distribution by using an elliptical multivariate $t$-distribution and the cumulative distribution function of the standard normal distribution as the skewing function. 

\begin{figure}[!ht]
\centering
\includegraphics[scale=1.0]{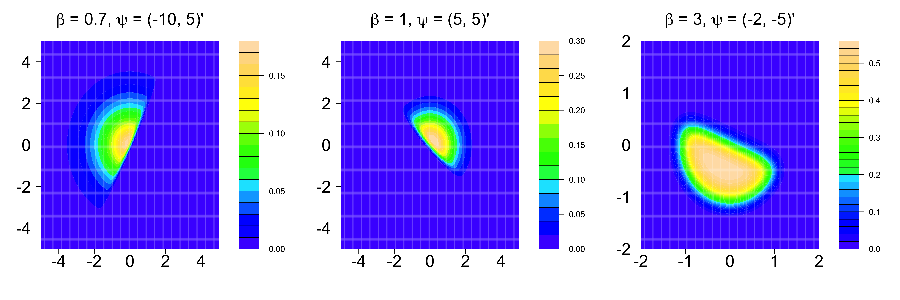}
\caption{Contours of the multivariate skew power exponential distribution for different values of the shape and skewness parameters with $\bmu=(0, 0)'$ and an identity scale matrix. The middle panel, with $\beta=1$, is a multivariate skew-normal distribution.}
\label{fig:MSPEcont}
\end{figure}

Herein, the skewness function is still the $N(0,1)$ cumulative distribution function while the elliptical distribution is now the MPE distribution. Specifically, a random vector $\bX$ follows a $p$-dimensional skew power exponential distribution if the density is of the form 
\begin{equation} \label{mspedef}
\begin{split}
f(\bx|\bmu,\bSigma,\beta, \bpsi)&=2 \, g(\bx|\bmu,\bSigma,\beta) \, \Phi(\bpsi' \bSigma ^{-1/2} (\bx - \bmu)),\\ 
&=\frac{2p\Gamma\left(\frac{p}{2}\right)}{\pi^{p/2}\Gamma\left(1+\frac{p}{2\beta}\right)2^{1+\frac{p}{2\beta}}} |\bSigma|^{-\frac{1}{2}} \exp\left\{ -\frac{1}{2} \delta(\bx)^\beta \right\} \Phi\left(\bpsi' \bSigma ^{-1/2} (\bx - \bmu)\right)
\end{split}\end{equation}
with location vector $\bmu$, scale matrix $\bSigma$, shape parameter $\beta$, and skewness vector $\bpsi$. Some special cases of this distribution include the skew-normal distribution $(\beta=1)$, a variant of a skew Laplace distribution $(\beta=0.5)$, the power exponential distribution $(\psi=0)$ and a generalization of the multivariate uniform distribution $(\beta\rightarrow\infty,\psi=0)$. Examples of contours of the MSPE distribution are given in Figure \ref{fig:MSPEcont}. 

\section{Mixtures of MSPE Distributions}
\label{sec:inference}
\subsection{Inference}
An iterative procedure is used for parameter estimation; specifically, a generalized expectation-maximization (GEM) algorithm \citep{dempster1977} with conditional maximization steps. The expectation-maximization (EM) algorithm \citep{dempster1977} is an iterative procedure in which the conditional expected value of the complete-data log-likelihood is maximized on each iteration to yield parameter updates. As opposed to the EM algorithm, the conditional maximization steps increase, rather than maximize, the conditional expected value of the complete-data log-likelihood in each iteration of a GEM algorithm.
Consider a random sample $\bx_1,\ldots,\bx_n$ from a $p$-dimensional MSPE mixture distribution from a population with $G$ subgroups. If we define 
$$z_{ig}=
\begin{cases}
1 & \text{if } \bx_{i} \text{ is from group } g,\\
0 & \text{otherwise},
\end{cases}$$
then the complete-data log-likelihood can be written as 
\begin{equation*}\begin{split}
\cL_c(\bTheta)  =\sum_{i=1}^{n} \sum_{g=1}^{G} z_{ig}\log&\Bigg[2 \pi_g \frac{p\Gamma\left(\frac{p}{2}\right)}{\Gamma\left(1+\frac{p}{2\beta_g}\right)2^{1+\frac{p}{2\beta_g}}\pi^{p/2}}  |\bSigma_g|^{-\frac{1}{2}}\\  
&\qquad\times\exp\left\{ -\frac{\delta_{ig}(\bx_i)^{\beta_g}}{2} \right\} \Phi\left(\bpsi'_{g} \bSigma_{g} ^{-1/2} (\bx_{i} - \bmu_{g})\right)\Bigg].
\end{split}\end{equation*}  

For parsimony, an eigen-decomposition is commonly imposed on component scale matrices using the re-parameterization $\bSigma_g=\lambda_g {\bGamma_g} {\bDelta_g} {\bGamma_g}'$, where $\bDelta_g$ is a diagonal matrix with entries proportional to the eigenvalues of $\bSigma_g$ (with $|\boldsymbol{\Delta}_g|=1$), $\lambda_g$ is the associated constant of proportionality, and $\boldsymbol{\Gamma}_g$ is a $p\times p$ orthogonal matrix of the eigenvectors of~$\bSigma_{g}$ with entries ordered according to the eigenvalues \citep{banfield1993,celeux1995}. A subset of eight models was considered in \cite{dang15} including the most parsimonious (EII) and the fully unconstrained (VVV) models, \edit{along with a possible constraint on $\beta_g$}, for their family of mixture models using elliptical power exponential distributions (Table~\ref{tab:models}). Herein, we consider the same eight models to form a family of mixtures of skewed power exponential distributions. 
\begin{table*}[ht]
	\caption{Nomenclature, scale matrix structure, and the number of free scale parameters for the eigen-decomposed family of models.} \label{tab:models}
	\begin{tabular*}{1.0\textwidth}{@{\extracolsep{\fill}}lllllr}
		\toprule
		Model & $\lambda_g$ & $\bDelta_g$ & $\bGamma_g$ & $\boldsymbol{\Sigma}_g$ & Free Parameters \\
		\midrule
		EII & Equal    & Spherical & --            & $\lambda \boldsymbol{I}$           & 1\\
		VII & Variable & Spherical & --            & $\lambda_g \boldsymbol{I}$         & $G$\\[3mm]
		EEI & Equal    & Equal     & Axis-Aligned & $\lambda \boldsymbol{\Delta}$      & $p$\\
		VVI & Variable & Variable  & Axis-Aligned & $\lambda_g \boldsymbol{\Delta}_g$  & $Gp$\\[3mm]
		EEE & Equal    & Equal     & Equal        & $\lambda\boldsymbol{\Gamma}\boldsymbol{\Delta}\boldsymbol{\Gamma}'$  & $p\left(p+1\right)/2$\\
		EEV & Equal    & Equal     & Variable     & $\lambda\boldsymbol{\Gamma}_g\boldsymbol{\Delta}\boldsymbol{\Gamma}_g'$  & $Gp(p+1)/2 - (G-1)p$ \\
		VVE & Variable & Variable  & Equal        & $\lambda_g\boldsymbol{\Gamma}\boldsymbol{\Delta}_g\boldsymbol{\Gamma}'$  & $p(p+1)/2 + (G-1)p$ \\
		VVV & Variable & Variable  & Variable   & $\lambda_g \boldsymbol{\Gamma}_g \boldsymbol{\Delta}_g \boldsymbol{\Gamma}_g'$  & $Gp\left(p+1\right)/2$ \\
		\bottomrule
	\end{tabular*}
	\bigskip
\end{table*}

After initialization (Section \ref{sec:initialize}), the algorithm proceeds as follows.\\
{\bf E-Step}: In the E-step, the group membership estimates $\hat{z}_{ig}$ are updated using 
\begin{equation*} \label{estep}
\hat{z}_{ig} \colonequals  \EE_{\widehat\bTheta} [Z_{ig}|\bx_i] = \frac{\hat{\pi}_g f \left(\bx_i| \hat\bmu_g, \hat\bSigma_g,\hat\beta_g,\hat\bpsi_g\right)} {\sum_{j=1}^G \hat\pi_j f \left(\bx_i| \hat\bmu_{j}, \hat\bSigma_{j},\hat\beta_j,\hat\bpsi_j\right)},
\end{equation*} for $i=1,\ldots,n$ and $g=1,\ldots,G$.\\
{\bf M-Step}: The update for $\pi_{g}$ is $\hat{\pi}_{g} = {n_g}/{n},$ where $n_g=\sum_{i=1}^n\hat{z}_{ig}$. However, the updates for ${\bmu}_g$, ${\bSigma}_g$, ${\beta}_g$ and ${\bpsi}_g$ are not available in closed form. %
For estimating $\beta_g$, either a Newton-Raphson method or a root finding algorithm may be used and is identical to the estimate in \cite{dang15}. In our implemented code, we constrain $\beta_{g}$ to be less than 20 for numerical stability. 
Let \begin{equation*} \label{estep}
\mathcal{Q} \colonequals  \EE_{\bTheta} [\cL_c(\bTheta|\bx)].
\end{equation*}
Then, a Newton-Raphson update is used for the location parameter $\hat{\bmu}_g$ with the following:
\begin{align} 
\frac{\partial \mathcal{Q}}{\partial \bmu_g} =& \hat{\beta}_g\sum_{i=1}^{n} \hat{z}_{ig} \delta_{ig}(\bx_i)^{\hat{\beta}_g-1} \hat{\bSigma}_g^{-1} (\bx_i-\hat{\bmu}_g) -  \sum_{i=1}^{n} \hat{z}_{ig} \frac{\phi(\bpsi'_g \bSigma_g ^{-1/2} (\bx_i-\hat{\bmu}_g))}{\Phi(\bpsi'_g \bSigma_g ^{-1/2} (\bx_i-\hat{\bmu}_g))} \bSigma_g ^{-1/2} \bpsi_g,\\
\frac{\partial^2 \mathcal{Q}}{\partial \bmu_g\bmu_g'} =& \hat{\beta}_g\sum_{i=1}^{n} \hat{z}_{ig} \left[-\delta_{ig}(\bx_i)^{\hat{\beta}_g-1} \hat{\bSigma}_g^{-1} 
 - 2(\hat{\beta}_g-1)\delta_{ig}(\bx_i)^{\hat{\beta}_g-2}\hat{\bSigma}_g^{-1} (\bx_i-\hat{\bmu}_g)(\bx_i-\hat{\bmu}_g)'\hat{\bSigma}_g^{-1} \right] \nonumber\\ \nonumber
& - \sum_{i=1}^{n} \hat{z}_{ig} \bpsi'_g \bSigma_g ^{-1/2} (\bx_i-\hat{\bmu}_g) \frac{\phi(\bpsi'_g \bSigma_g ^{-1/2} (\bx_i-\hat{\bmu}_g))}{\Phi(\bpsi'_g \bSigma_g ^{-1/2} (\bx_i-\hat{\bmu}_g))} \hat{\bSigma}_g^{-1} \bpsi_g\bpsi'_g\hat{\bSigma}_g^{-1}  \\ \nonumber
& - \sum_{i=1}^{n} \hat{z}_{ig} \left[\frac{\phi(\bpsi'_g \bSigma_g ^{-1/2} (\bx_i-\hat{\bmu}_g))}{\Phi(\bpsi'_g \bSigma_g ^{-1/2} (\bx_i-\hat{\bmu}_g))}\right]^2 \hat{\bSigma}_g^{-1} \bpsi_g\bpsi'_g\hat{\bSigma}_g^{-1},\nonumber
\end{align}
where $\delta_{ig}(\bx_i)\colonequals%
\left (\bx_i-\hat{\bmu}_g \right)'\hat{\bSigma}_g^{-1} \left(\bx_i-\hat{\bmu}_g \right)$.

For estimating the skewness parameter $\bpsi$, the density is first re-parameterized as 
\begin{align*}
f(\bx|\bmu,\bSigma,\beta, \bpsi)=&2 \, g(\bx|\bmu,\bSigma,\beta) \, \Phi(\bpsi' \bSigma ^{-1/2} (\bx - \bmu)) \\ \nonumber
=&2 \, g(\bx|\bmu,\bSigma,\beta) \, \Phi(\boeta' (\bx - \bmu)), \nonumber
\end{align*}
where $\boeta = \bSigma^{{-1/2}}\bpsi$. A quadratic lower-bound principle \citep{bohning1988,hunter2004} on the relevant part of the complete-data log-likelihood using the re-parameterized density uses the following property to construct a quadratic minorizer:
\begin{equation*}
\log(\Phi(s)) \geq \log(\Phi(s_0)) + \frac{\phi(s_0)}{\Phi(s_0)} (s - s_0) + \frac{1}{2} (-1) (s - s_0)^2,
\end{equation*}
where $-1$ is the lower bound of the second derivative in the Taylor series around $s_{0}$. Then, an estimate for $\boeta_{g}$ yields
\begin{align*}
\boeta_{g} = \left[ \sum_{i=1}^{n} \hat{z}_{ig} (\bx_i-\hat{\bmu}_g) (\bx_i-\hat{\bmu}_g)'\right]^{{-1}} &\left[\sum_{i=1}^{n} \hat{z}_{ig} \frac{\phi(\boeta'_{g0} (\bx_i-\hat{\bmu}_g))}{\Phi(\boeta'_{g0} (\bx_i-\hat{\bmu}_g))} (\bx_i-\hat{\bmu}_g) \right. \\
 &\qquad \left. + \sum_{i=1}^{n} \hat{z}_{ig} (\bx_i-\hat{\bmu}_g)(\bx_i-\hat{\bmu}_g)' \boeta_{g0}\right]
\end{align*}
and we can back-transform to obtain $\bpsi_g = \bSigma^{1/2}_g\boeta_g$. 

For the scale matrices $\bSigma_g$, the estimation makes use of minorization-maximization algorithms \citep{hunter2000,hunter2004} by exploiting the concavity of the functions containing $\bSigma_g$ (or parts of its decomposition) and accelerated line search algorithms on the Stiefel manifold \citep{absil2009,browne2014}, with different updates depending on whether the latest estimate for $\beta_g$ is less than 1 or is greater than or equal to 1. For more details, see \cite{dang15}. Combining the constraints of the eigen-decomposition in Table~\ref{tab:models}, with constraining $\beta_g$ to be equal or different between groups, results in a family of 16 models. For example, a VVIE model represents a VVI scale structure (as in Table \ref{tab:models}) and the shape parameter constrained to be equal between groups ($\beta_g=\beta$).

\subsection{Initialization}
\label{sec:initialize}
It is well known that the performance of the EM algorithm depends heavily on the starting values. The following strategy is adopted. The group memberships are initialized using \edit{a combination of emEM approach \citep{biernacki2003}, $k$-means algorithm \citep{hartigan79}, and random soft starts. Specifically, the most superior run (highest log-likelihood) from 10 different $k$-means-based \textit{short EM} runs is chosen for a \textit{long EM} run. This process is repeated with random soft starts. The fits of the two \textit{long EM} runs are compared based on a model selection criterion (see Section \ref{modelselection}) to choose a best model. Once these initial memberships are set, the $\bmu_g$ and $\bSigma_g$ are initialized using a constrained model.} 
The kurtosis parameters $\beta_g$ are initialized to 0.5, and the skewness is initialized as a zero vector.

\subsection{Convergence, Model Selection, and Performance Assessment} \label{modelselection}
Following \cite{lindsay1995} and \cite{mcnicholas2010a}, the iterative GEM algorithm is stopped based on the Aitken's acceleration \citep{aitken1926}. Specifically, an asymptotic estimate of the log-likelihood at iteration $k+1$ is compared with the current log-likelihood value and considered converged when the difference is less than some positive $\epsilon$. We use $\epsilon=0.005$ for the analyses herein.

In a general clustering scenario, the number of groups is generally not known a~priori and the covariance model is not known. Therefore, a model selection criterion is required. The most common criterion for model selection is the Bayesian information criterion \cite[BIC;][]{schwarz78}. The BIC can be written as
\begin{equation}\label{eqn:bic}
\text{BIC}=2l(\hat\bvartheta)-m\log n,
\end{equation}
where $m$ is the number of free parameters, $n$ is the sample size and $l(\hat\bvartheta)$ is the maximized log-likelihood. When written as in \eqref{eqn:bic}, a greater BIC represents a superior model fit. The integrated complete likelihood \cite[ICL;][]{biernacki00} was also considered for model selection; however, in initial testing, the ICL did not consistently outperform the BIC in simulations and thus, for the remainder of this manuscript, we use only the BIC.

To evaluate classification performance, we use the adjusted Rand index \cite[ARI;][]{hubert85}. The ARI compares two different partitions; specifically, in our case, the estimated classification and the (known) true classifications. The ARI takes a value of 1 when there is perfect classification and has expected value 0 under random classification. See 
\cite{steinley04} for extensive details on the ARI.

\section{Analyses}
\label{sec:results}

\subsection{Overview}
\edit{The performance of the MSPE mixture models is compared with mixture model implementations based on the MPE distribution \citep{dang15}, as well as implementations from the {\tt mixture} package \citep{mixturefromR} of the generalized hyperbolic distribution (\texttt{ghpcm}) and the Gaussian distribution (\texttt{gpcm}).} We chose these mixtures for comparison as Gaussian mixtures remain widely used and the generalized hyperbolic distribution has special cases that include some parameterizations of inverse Gaussian, variance-gamma, skew-$t$ (note there are formulations of the skew-$t$ distribution that cannot be obtained from the generalized hyperbolic), multivariate normal-inverse Gaussian, and asymmetric Laplace distribution. Using these comparators, we obtain comparisons to mixtures based on purely elliptical (Gaussian), elliptical with flexible kurtosis modeling (MPE), and skewed (generalized hyperbolic) distributions. \edit{For a fair comparison, we restrict the models in the other implementations to those in Table~\ref{tab:models}.} In addition, we use BIC as the model selection criterion, we run $G = 1,\ldots,4$ for all simulations and real data analyses, and \edit{we use the same starting soft memberships for all comparator models}. Data from the MSPE distribution is simulated using a Metropolis-Hastings rule.

\subsection{Simulations}
\subsubsection{Simulation 1: Heavy and light-tailed skewed clusters}

\begin{figure}[!ht]
\centering
\includegraphics[scale=0.90,angle=0]{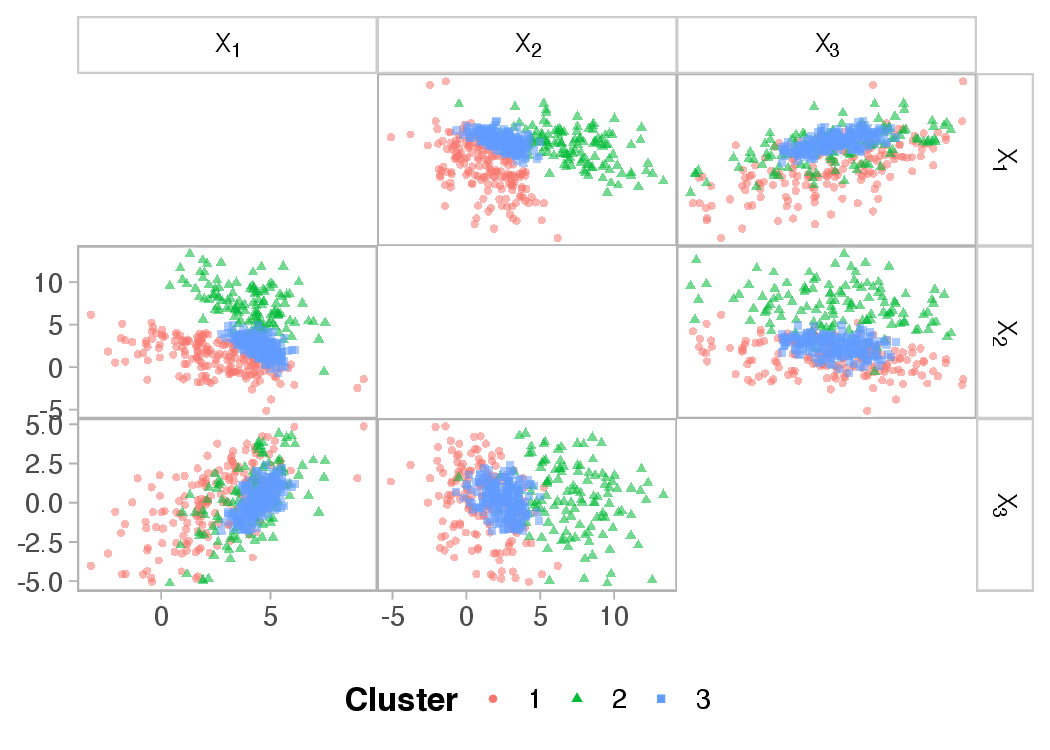}
\caption{An example scatterplot matrix of the three-component mixture of Simulation 1.}
\label{fig:sim1}
\end{figure}

A three-component mixture is simulated with 500 observations in total. Group sample sizes are sampled from a multinomial distribution with mixing proportions $(0.35, 0.25, 0.4)'$. The first component is simulated from a heavy-tailed three-dimensional MSPE distribution with $\bmu_1 = (3, 3, 0)'$, $\beta_1=0.85$, and $\bpsi_1=(-5, -10, 0)'$. The second component is  simulated with $\bmu_2 = (3, 6, 0)'$, $\beta_2=0.9$, and $\bpsi_2=(15, 10, 0)'$. The third component is simulated with light tails with $\bmu_3 = (4, 2, 0)'$, $\beta_3=2$, and $\bpsi_3=\bpsi_2$. Lastly, the scale matrices were common to all three components with $\text{diag}(\boldsymbol{\Delta}_{g}) = (4, 3, 1)'$ and 
$$\boldsymbol{\Gamma}_g=\begin{pmatrix}
0.36 & 0.48 & -0.80 \\ 
-0.80 & 0.60 & 0.0 \\ 
0.48 & 0.64 & 0.6 \\ 
 \end{pmatrix},$$
for $g=1,2,3$.
The simulated components are not well separated (an example scatterplot matrix is given in Figure~\ref{fig:sim1}). All four mixture implementations are run on 100 such datasets for $G=1,\ldots,4$.

For the MSPE family, a three-component \edit{(four-component) model is selected by the BIC 98 (2) times}. For the MPE family, the BIC selects a three-component (four-component) model \edit{87 (13) times}. When the four-component models are selected, typically, this is because the model chosen has split up the heavy-tailed cluster into two separate components. \edit{The exception is that for two chosen four-component MPE solutions, the light-tailed cluster was split into two separate components.}
For the {\tt gpcm} family, the BIC selects a three-component \edit{(four-component) model \edit{99 (1) times}}. On the other hand, for the {\tt ghpcm} algorithm, \edit{one-component, two-component, and three-component models are selected 3, 46, and 51 times, respectively}. This under-fitting --- \edit{the heavy-tailed and light-tailed components} are merged --- may be due to the use of the BIC. 

The ARI values for the selected MSPE models range from \edit{0.71} to 0.91, with a median (mean) ARI value of \edit{0.83 (0.83)}. The selected MPE models yield ARI values ranging between \edit{0.67} and 0.91, with a median (mean) value of 0.80 (0.79). The selected {\tt gpcm} models yield ARI values ranging between 0.71 and 0.91, with a median (mean) value of 0.80 (0.80). Similarly, the {\tt ghpcm} algorithm yields ARI values ranging between \edit{0 and 0.82}, with a median (mean) value of 0.41 (0.47). For the MSPE models, an EEEV model was selected \edit{68 out of the 100 times, with a less parsimonious model selected the other times. To demonstrate parameter recovery, a similar simulation setup was carried out and findings are provided in the Appendix}. 

\subsubsection{Simulation 2: Heavy-tailed and Gaussian skewed clusters}

A three-component mixture is simulated with 500 observations in total. Group sample sizes are sampled from a multinomial distribution with mixing proportions $(0.3, 0.45, 0.25)'$. The first component is simulated from a three-dimensional skewed normal distribution (i.e., $\beta_1=1$) with $\bmu_1 = (0, 1, 2)'$ and $\bpsi_1=(3, 5, 10)'$. The second component is simulated from a heavy-tailed three-dimensional MSPE distribution with $\bmu_2 = (0, 4, 2)'$, $\beta_2=0.8$, and $\bpsi_2=(-3, 5, -5)'$. The third component is simulated with $\bmu_3 = (-2, -3, 0)'$, $\beta_3=0.9$, and $\bpsi_3=(5, 10, -5)'$. 
Lastly, the scale matrices are
$$\bSigma_1=\begin{pmatrix}
1.00 & 0.50 & 0.40 \\ 
0.50 & 1.50 & 0.35 \\ 
0.40 & 0.35 & 1.20 \\ 
 \end{pmatrix}
\qquad \text{and} \qquad 
\bSigma_2=\bSigma_3=\begin{pmatrix}
1.00 & 0.30 & 0.20 \\ 
0.30 & 1.50 & 0.30 \\ 
0.20 & 0.30 & 1.20 \\ 
 \end{pmatrix}.$$
Again, the simulated components are not well separated and all four mixture implementations are run on 100 such datasets for $G=1,\ldots, 4$. 

For the MSPE family, a three-component (four-component) component model is selected by the BIC \edit{95 (5)} times. For the MPE mixture, the BIC selected a \edit{three-component (four-component) model 99 (1) times}. Interestingly, for the {\tt ghpcm} mixtures, the BIC selects a three-component model all 100 times. On the other hand,  for the {\tt gpcm} family, the BIC selects a three-component (four-component) model \edit{94 (6) times}. In all cases when the four-component models are selected, this is because the model chosen has split up one of the heavy-tailed clusters into two components. 

The ARI values for the selected MSPE models range from \edit{0.74 to 0.98}, with a median (mean) ARI value of 0.94 (0.93). The selected MPE models yield ARI values ranging between \edit{0.72} and 0.96, with a median (mean) value of 0.90 \edit{(0.89)}. Similarly, the {\tt ghpcm} algorithm yields ARI values ranging between \edit{0.70 and 0.94, with a median (mean) value of 0.86 (0.85)}. The selected {\tt gpcm} models yield ARI values ranging between \edit{0.68} and 0.95, with a median (mean) value of 0.90 (0.88). For the MSPE models, an EEEV model is selected 76 out of 100 times.

\subsubsection{Simulation 3: Two light-tailed elliptical clusters}
A simulation from \cite{dang15} is replicated, where a two-component EIIV model is simulated with 450 observations with the sample sizes for each group sampled from a binomial distribution with success probability $0.45$. Both components had identity scale matrices and zero skewness. The first component is simulated from a two-dimensional MPE distribution with $\bmu_1 = (0,0)'$ and $\beta_1=2$ while the second component is simulated using $\bmu_2 = (2,0)'$ and $\beta_2=5$. Again, the simulated components are not well separated. All four algorithms are run on 100 such datasets. 
For the MSPE and MPE families, a two-component model is selected by the BIC for 100 and 99 datasets, respectively. The dataset where a three-component model is selected for the MPE models involves a cluster of \edit{four} observations that are tightly clustered with tiny eigenvalues. On the other hand, for the {\tt gpcm} family, the BIC selects a two-, three-, and four-component model \edit{80, 12, and 8 times}, respectively. Similarly, for the {\tt ghpcm} algorithm, one-, two-, and three-component models are selected \edit{1, 95, and 4 times}, respectively. Here, more components are being fitted to deal with the light-tailed nature of the data.

For the MPE family (which was used to simulate the data), the ARI values for the selected models range from 0.81 to 0.95, with a median (mean) ARI value of 0.88 (0.88). The MSPE family performed similarly, as expected --- the MPE is a special case of the MSPE family --- with the ARI values for the selected models ranging from 0.81 to 0.94, with a median (mean) ARI value of 0.88 (0.88). The selected {\tt gpcm} models yield ARI values ranging between 0.35 and 0.96, with a median (mean) value of 0.85 (0.80). Similarly, the {\tt {\tt ghpcm}} algorithm yields ARI values ranging between 0 and \edit{0.91}, with a median (mean) value of \edit{0.81 (0.80)}. For the MSPE and MPE models, an EIIV model was selected 89 and 95 times out of 100, respectively. Because the ranges of ARI can be more reflective of one poor or great fit, the median, first and third quartile of the ARIs for the selected models are summarized in Table \ref{tab:simulation_summary} for all simulations.
\begin{table}[ht]
\centering
\caption{Performance comparison of four mixture model families on simulations. For each simulation and implementation, a frequency table of number of groups of the best selected model according to the BIC is provided. Median ARI is provided as well as the first and third quartiles for ARI across 100 datasets in each simulation in parenthesis as table entries.}
\label{tab:simulation_summary}
\begin{tabular}{@{}lllll@{}}
\cmidrule(r){1-5}
     &             & Simulation 1      & Simulation 2      & Simulation 3          \\ \cmidrule(r){1-5}
MSPE  & Frequencies & \edit{3 (98); 4 (2)}            & 3 \edit{(95); 4 (5)}     & 2 (100)               \\
 & ARI & \textbf{\edit{0.83 (0.81, 0.86)}} & \textbf{0.94 (\edit{0.93}, 0.95)} & \textbf{0.88 (\edit{0.85}, 0.90)} \\ \\
MPE   & Frequencies & \edit{3 (87); 4 (13)}     & \edit3 {(99); 4 (1)}          & 2 (99); 3 (1)         \\
 & ARI & 0.80 (0.76, 0.83) & 0.90 (0.88, 0.91)          & \textbf{0.88 (0.86, 0.90)} \\ \\
{\tt ghpcm}  & Frequencies & \edit{1 (3); 2 (46); 3 (51)}     & 3 (100)           & \edit{1 (1); 2 (95); 3 (4)} \\
     & ARI         & \edit{0.41 (0.30, 0.66)} & \edit{0.86 (0.82, 0.88)} & \edit{0.81 (0.79, 0.85})     \\ \\
{\tt gpcm} & Frequencies & \edit{3 (99); 4 (1)}            & \edit{3 (94); 4 (6)}     & \edit{2 (80); 3 (12); 4 (8)} \\
     & ARI         & 0.80 (0.77, 0.82) & 0.90 (0.87, 0.91) & 0.85 (\edit{0.77}, 0.89)     \\ \cmidrule(r){1-5}
\end{tabular}
\end{table}

\subsection{Dataset Descriptions}
For assessment of performance on real data, we considered the following ``benchmark'' datasets, i.e., datasets often used for comparison of clustering algorithms, \edit{of various sizes and dimensionalities} available through various {\sf R} packages:
\begin{itemize}
\item \edit{ \textbf{Body Dataset}: The {\tt body} data from the {\tt gclus} package \citep{gclus} has 24 measurements on body dimension, age, etc., for 507 individuals (247 men and 260 women).}
\item \edit{\textbf{Coffee Dataset}: The {\tt coffee} dataset \citep{streuli1973}, obtained from the {\tt pgmm} package \citep{pgmm}, has 12 chemical measurements on two types of coffee (\textit{arabica} and \textit{robusta}).}
\item \textbf{Female Voles Dataset}: The {\tt fvoles} dataset \citep{flury1997} has measurements on six size variables and age of 86 female voles from two different species (\textit{californicus} and \textit{ochrogaster}).
\item \edit{\textbf{Olive Dataset}: The {\tt olive} dataset \citep{forina1982}, obtained from the {\tt pgmm} package, has percentage composition of eight fatty acids of olive oils from three regions of Italy.}
\item \edit{\textbf{Penguins Dataset}: The {\tt penguins} dataset \citep{penguins} has measurements  on bill length, bill depth, flipper length, and body mass of 342 penguins from three different species (\textit{Adélie}, \textit{Chinstrap}, and \textit{Gentoo}).}
\item \textbf{Wine Dataset}: The thirteen variable {\tt wine} dataset, obtained from \cite{gclus}, has 13 different measurements of chemical aspects of 178 Italian wines from three different types of wine. 
\end{itemize}
Several other datasets were also considered, on which performance of the different algorithms is summarized in the Appendix.

\subsection{Unsupervised Classification} \label{sec:unsup}
Unsupervised classification is performed on (scaled) datasets mentioned above using the same comparison distributions as on the simulated data. 
The ARI and the number of groups chosen by the BIC is shown in Table~\ref{tab:unsup} along with the classification tables in Table~\ref{tab:classtab}. %
\begin{table}[ht]
	\centering 
	\caption{Performance comparison of four mixture model families on real data for the unsupervised scenarios. Sample size, dimensionality, and the number of known groups (i.e., classes) are in parenthesis following each dataset name. For each implementation, the ARI and the number of selected components as well as the BIC are provided in parenthesis.}
	\label{tab:unsup}
\resizebox{\textwidth}{!}{%
\begin{tabular}{p{6.3cm}llll}
\hline
Data & MSPE & MPE & {\tt ghpcm} & {\tt gpcm}\\

\hline

Body ($n=507, p=24, G=2$) & \textbf{0.94} (2; $-18681$) & 0.59 (4; $-18670$) & 0.42 (4; $-17340$) & 0.48 (4; $-18842$)\\
Coffee ($n=43, p=12, G=2$) & \textbf{1} (2; $-1373$) & 0.38 (3; $-1297$) & 0.27 (4; $-497$) & 0.38 (3; $-1311$)\\

Female voles ($n=86, p=7, G=2$) & \textbf{0.91} (2; $-1327$) & 0.8 (3; $-1304$) & 0.82 (2; $-1350$) & \textbf{0.91} (2; $-1317$)\\
Olive ($n=572, p=8, G=3$) & \textbf{0.7} (4; $-5271$) & 0.69 (4; $-5297$) & \textbf{0.7} (4; $-5230$) & 0.67 (4; $-5437$)\\
Penguins ($n=342, p=4, G=3$) & \textbf{0.96} (3; $-2532$) & 0.81 (4; $-2502$) & 0.95 (3; $-2449$) & 0.76 (4; $-2497$)\\
Wine ($n=178, p=13, G=3$) & \textbf{0.92} (3; $-5474$) & 0.68 (4; $-5438$) & 0.8 (4; $-4833$) & 0.8 (4; $-5376$)\\
\hline
\end{tabular}
}
\end{table}

\edit{In the case of the {\tt body} dataset, heavy-tailed components with some skewness are fit with the best selected MSPE model (an EEEV model). For the other three comparators, the BIC selected a four-component model, with the fewest misclassifications for the MPE model. The {\tt coffee} dataset is a small dataset ($n=43$), on which the selected MSPE model (heavy-tailed) had perfect classification. The other three comparators overfit the number of components, with MPE and {\tt gpcm} obtaining the same classification and the {\tt ghpcm} model fitting two small clusters containing six and seven observations, respectively. In the case of the {\tt fvoles} dataset, skewed yet light-tailed clusters are fit. However, the {\tt gpcm} model performed equally well here while the {\tt ghpcm} model misclassified two more observations in comparison. On the other hand, the selected MPE model fit three components (with one component containing only six observations). In the case of the {\tt olive} data, the BIC selected a similarly performing, four-component, solution for all four comparators ($G=3$) with olive oils from Southern Italy split into two components. The MSPE model fit heavy-tailed components. On the {\tt penguins} data, the two skewed mixtures, MSPE and {\tt ghpcm} performed best, fitting the `correct' number of components. The MSPE model fit heavy (with some skewness) clusters. The selected models from the MPE and {\tt gpcm} split different species into two different components with the best fit {\tt gpcm} model having the largest number of misclassifications of all the comparators. On the other hand, in the case of the {\tt wine} data, the MSPE mixture fit heavy-tailed components with five misclassifications. The BIC selected a four-component solution for the other three comparators. }

\edit{Note that, for the {\tt body} and {\tt wine} datasets, the findings above differ from those previously reported in \cite{dang15} for the MPE mixtures; this is likely due to different starting values.} For example, in \cite{dang15}, a solution was obtained using MPE mixtures which fit three components and misclassified only one observation; however, here the selected MPE mixture based on the BIC was a four-component model, likely due to different starting values (one of the MPE mixture fits also only had one misclassification; this was the third best model according to the BIC). Note that in the semi-supervised case as well (Table~\ref{tab:semisup}), the MSPE mixtures perform the best followed by the MPE mixtures. The estimates of the $\beta_g$ parameters from the MSPE fit indicate heavy tails while {\tt ghpcm} overfits the number of components.

It is interesting to note that while the BIC is used to select one model from a family of models, there is no guarantee that this model, whether within a family or between families, provides the most superior clustering performance.
\begin{table}[ht]
\centering
\caption{Classification tables for the cluster analysis of each model and dataset.}
\scalebox{0.74}{
\begin{tabular}{ccccc}
\hline
Dataset& MSPE & MPE & {\tt ghpcm} & {\tt gpcm}\\
\hline
\\
{\tt body}&\begin{tabular}{rrr}
  \hline
 & 1 & 2 \\ 
  \hline
Women &  256 &   4 \\ 
  Men &   4 &  243 \\ 
   \hline
\end{tabular}&\begin{tabular}{rrrr}
\hline
1 & 2 & 3 & 4\\
\hline
200 & 0 & 4 & 56\\
1 & 167 & 77 & 2\\
\hline
\end{tabular}&\begin{tabular}{rrrr}
\hline
1 & 2 & 3 & 4\\
\hline
133 & 9 & 118 & 0\\
10 & 117 & 2 & 118\\\hline
\end{tabular}&\begin{tabular}{rrrr}
\hline
1 & 2 & 3 & 4\\
\hline
148 & 0 & 109 & 3\\
0 & 142 & 5 & 100\\
\hline
\end{tabular}\\ \\
{\tt coffee}&\begin{tabular}{lrr}
  \hline
 & 1 & 2 \\ 
  \hline
Arabica &  36 &   0 \\ 
Robusta &   0 &  7 \\ 
   \hline
\end{tabular}&\begin{tabular}{rrr}
\hline
1 & 2 & 3\\
\hline
22 & 14 & 0\\
0 & 0 & 7\\\hline
\end{tabular}&\begin{tabular}{rrrr}
\hline
1 & 2 & 3 & 4\\
\hline
19 & 11 & 0 & 6\\
0 & 0 & 7 & 0\\\hline
\end{tabular}&\begin{tabular}{rrr}
\hline
1 & 2 & 3\\
\hline
22 & 14 & 0\\
0 & 0 & 7\\\hline
\end{tabular}\\ \\
{\tt fvoles}&\begin{tabular}{rrr}
  \hline
 & 1 & 2 \\ 
  \hline
{\it californicus} &  41 &   0 \\ 
{\it  ochrogaster} &   2 &  43 \\ 
   \hline
\end{tabular}&\begin{tabular}{rrr}
\hline
1 & 2 & 3\\
\hline
35 & 0 & 6\\
2 & 43 & 0\\
\hline
\end{tabular}&\begin{tabular}{rr}
\hline
1 & 2\\
\hline
38 & 3\\
1 & 44\\
\hline
\end{tabular}&\begin{tabular}{rr}
\hline
1 & 2\\
\hline
41 &   0 \\ 
2 & 43\\
\hline
\end{tabular}\\ \\
%
%
{\tt olive}&\begin{tabular}{lrrrr}
\hline
& 1 & 2 & 3 & 4\\
\hline
Southern Italy&217 & 0 & 106 & 0\\
Sardinia&0 & 0 & 0 & 98\\
Northern Italy&0 & 151 & 0 & 0\\
\hline
\end{tabular}&\begin{tabular}{rrrr}
\hline
1 & 2 & 3 & 4\\
\hline
217 & 0 & 106 & 0\\
0 & 0 & 0 & 98\\
0 & 150 & 0 & 1\\\hline
\end{tabular}&\begin{tabular}{rrrr}
\hline
1 & 2 & 3 & 4\\
\hline
222 & 0 & 101 & 0\\
0 & 0 & 0 & 98\\
0 & 149 & 0 & 2\\
\hline
\end{tabular}&\begin{tabular}{rrrr}
\hline
1 & 2 & 3 & 4\\
\hline
196 & 0 & 127 & 0\\
0 & 0 & 0 & 98\\
0 & 151 & 0 & 0\\
\hline
\end{tabular}\\ \\

{\tt penguins}&\begin{tabular}{lrrr}
\hline
  & 1 & 2 & 3\\
\hline
{\it Adelie}&149 & 0 & 2\\
{\it Chinstrap}&3 & 0 & 65\\
{\it Gentoo}&0 & 123 & 0\\\hline
\end{tabular}&\begin{tabular}{rrrr}
\hline
1 & 2 & 3 & 4\\
\hline
150 & 1 & 0 & 0\\
4 & 64 & 0 & 0\\
0 & 0 & 63 & 60\\
\hline
\end{tabular}
&\begin{tabular}{rrrr}
  \hline
1 & 2 & 3 \\ 
  \hline
151 & 0 & 0\\
6 & 0 & 62\\
0 & 123 & 0\\
   \hline
\end{tabular} &\begin{tabular}{rrrr}
\hline
1 & 2 & 3 & 4\\
\hline
0 & 96 & 1 & 54\\
0 & 5 & 63 & 0\\
123 & 0 & 0 & 0\\
\hline
\end{tabular}\\ \\

{\tt wine}&\begin{tabular}{rrrr}
  \hline
 & 1 & 2 & 3 \\ 
  \hline
Barolo&0 & 59 & 0\\
Grignolino&66 & 0 & 5\\
Barbera&0 & 0 & 48\\   \hline
\end{tabular}&\begin{tabular}{rrrr}
  \hline
1 & 2 & 3 & 4 \\ 
  \hline
0 & 50 & 0 & 9\\
5 & 0 & 46 & 20\\
48 & 0 & 0 & 0\\
   \hline
\end{tabular}&\begin{tabular}{rrrr}
  \hline
1 &  2 & 3 & 4\\ 
  \hline
57 & 0 & 0 & 2\\
0 & 1 & 47 & 23\\
0 & 48 & 0 & 0\\
   \hline
\end{tabular}
&\begin{tabular}{rrrr}
  \hline
1 &  2 & 3 & 4\\ 
  \hline
0 & 54 & 0 & 5\\
57 & 0 & 1 & 13\\
0 & 0 & 46 & 2\\
   \hline
\end{tabular}\\ \\
\hline
\end{tabular}}
\label{tab:classtab}
\end{table}

\subsection{Semi-Supervised Classification}
Using the same datasets as in the unsupervised classification case, semi-supervised classification is now considered. Following \cite{mcnicholas10c}, the (observed) likelihood in the model-based classification framework can be written
\begin{equation*}
\mathcal{L}(\boldsymbol{\vartheta} \mid \bx_1, \ldots, \bx_n)=\prod_{i=1}^{k} \prod_{g=1}^G [\pi_g f(\bx_{i}|\btheta_g)]^{z_{ig}}
\prod_{j=k+1}^{n} \sum_{h=1}^G [\pi_h f(\bx_{j}|\btheta_h)],
\end{equation*}
where the first $k$ observations have known component memberships (i.e., labels), $f(\bx_{i}|\btheta_g)$ is the $g$th component density, and $\pi_g$ and $z_{ig}$ have the usual interpretations.

For each dataset previously considered, we take 25 labelled/unlabelled splits with 25\% supervision. In Table \ref{tab:semisup}, we display the median ARI values along with the first and third quartiles over the 25 splits. %
\edit{Generally, and as one would expect, performance in the semi-supervised scenarios was found to be better than in the fully unsupervised scenarios. 
In the case of the {\tt wine} data, as mentioned in Section \ref{sec:unsup}, the MSPE mixtures clearly exhibit the best performance (perfect classification), followed by the MPE mixtures. 
On the {\tt fvoles}, {\tt olive}, and {\tt penguins} datasets, all comparators performed similarly. For the {\tt penguins} dataset, the highest ARI was achieved by the two power exponential based mixtures (only one misclassification based on 25\% supervision). On the other hand, on the {\tt coffee} dataset, all models performed well although {\tt gpcm} and {\tt ghpcm} had some runs with poor fits.
On the {\tt body} dataset, the two power exponential-based mixtures performed best, with the highest ARI achieved with three misclassifications overall (75\% unlabelled), followed by {\tt gpcm}. 
}
\begin{table}[ht]
	\centering
	\caption{Median ARI values along with first and third quartiles in parenthesis for the four different models for each dataset for the semi-supervised runs. Size, dimensionality and the number of known groups (i.e., classes) are in parenthesis following each dataset name.}
	\label{tab:semisup}
	\resizebox{\textwidth}{!}{%
\begin{tabular}{p{6.4cm}llll}
\hline
Data & MSPE & MPE & ghpcm & gpcm\\
\hline

Body ($n=507, p=24, G=2$) & \textbf{0.94 (0.93, 0.95)} & \textbf{0.94 (0.93, 0.95)} & 0 (0, 0.68) & 0.92 (0.9, 0.94)\\
Coffee ($n=43, p=12, G=2$) & \textbf{1 (1, 1)} & \textbf{1 (1, 1)} & 1 (0, 1) & 1 (0, 1)\\

Female voles ($n=86, p=7, G=2$) & \textbf{0.94 (0.88, 0.94)} & \textbf{0.94 (0.88, 0.94)} & \textbf{0.94 (0.88, 0.94)} & \textbf{0.94 (0.88, 0.94)}\\

Olive ($n=572, p=8, G=3$) & 1 (0.99, 1) & \textbf{1 (1, 1)} & 0.99 (0.98, 0.99) & \textbf{1 (1, 1)}\\
Penguins ($n=342, p=4, G=3$) & \textbf{0.96 (0.95, 0.97)} & \textbf{0.96 (0.95, 0.97)} & \textbf{0.96 (0.95, 0.97)} & \textbf{0.96 (0.95, 0.97)}\\

Wine ($n=178, p=13, G=3$) & \textbf{1 (1, 1)} & 0.98 (0.98, 0.98) & 0.8 (0, 0.85) & 0.91 (0.83, 0.91)\\
\hline
\end{tabular}	} 
\end{table}

\section{Discussion}
\label{sec:discussion}
A multivariate skewed power exponential distribution is introduced that is well suited for density estimation purposes for a wide range of data with non-Gaussian clusters. The family of 16 MSPE mixtures presented herein allow for robust mixture models for model-based clustering on skewed as well as symmetric components. These models can model components with varying levels of peakedness and tail-weight (light, heavy, Gaussian) simultaneously with skewness. As a result, these models are well suited to model heterogeneous data with non-Gaussian components.

In addition to these properties, special cases of the MSPE distribution include the skew-normal distribution among others. The performance of such mixtures for clustering is investigated on a wide range of simulated scenarios --- on heavy-tailed, light-tailed, Gaussian, and skewed components, and combinations thereof --- and on real data \edit{of varying dimensionalities and sizes} commonly used for illustrating clustering and classification. At present, model selection is performed using the BIC and, although this performs well in most cases, it is by no means perfect and alternative criteria could be considered in more detail. \edit{Standard errors of parameter estimates were not considered herein; this could be implemented via the standard information-based method to obtain the asymptotic covariance matrix of the estimates or via bootstrapping \citep{basford1997, lin14}}. 

Through simulations, we showed scenarios where such skewed mixtures are comparative to or better than widely used elliptical mixture models (mixtures of Gaussians) or skewed mixture models gaining increasing attention (mixtures of generalized hyperbolic distributions). When looking at real datasets, we compared in the context of both unsupervised classification (i.e., clustering) and semi-supervised classification. On these, the MSPE model performed just as well or better on most of the investigated datasets compared to three other mixture model families/algorithms. %
The analysis on the real datasets in the unsupervised case displayed some possible weaknesses, which may be related to initializations or the choice of BIC as the model selection criterion. %
Performance improved substantially for some of the datasets when a small level of supervision was introduced. 

There are numerous areas of possible future work. One such area would be to consider a mixture of factor analyzers with the MSPE distribution for high dimensional data. A matrix variate extension, in a similar manner to \cite{gallaugher18}, might also be interesting for modeling three-way data.

\section*{Acknowledgements}
This research was supported by the Natural Sciences and Engineering Research Council of Canada through their Discovery Grants program for Dang, Browne, and McNicholas, the Banting Postdoctoral Fellowship for Gallaugher, as well as the Canada Research Chairs program for McNicholas.

\section*{Code and Data Accessibility Statement}
All data used here are publicly available; references have been provided within the bibliography. An implementation of mixtures of skewed power exponential distributions and mixtures of power exponential distributions is available as the {\sf R} package {\tt mixSPE} \citep{dang21}. 

\section*{Ethical declarations}
The authors declare no competing interests.


\appendix
\section{Appendix}

\setcounter{table}{0}
\renewcommand{\thetable}{A\arabic{table}}

\subsection{Parameter recovery}

\edit{A three-component mixture is simulated with 500 observations in total. Group sample sizes are sampled from a multinomial distribution with mixing proportions $(0.2, 0.34, 0.46)'$. The first component is simulated from a heavy-tailed three-dimensional MSPE distribution with $\bmu_1 = (12, 14, 0)'$, $\beta_1=0.85$, and $\bpsi_1=(-5, -10, 0)'$. The second component is  simulated with $\bmu_2 = (-3, -10, 0)'$, $\beta_2=0.9$, and $\bpsi_2=(15, 10, 0)'$. The third component is simulated with light tails with $\bmu_3 = (3, 1, 0)'$, $\beta_3=2$, and $\bpsi_3=\bpsi_2$. Lastly, the scale matrices were common to all three components with $\text{diag}(\boldsymbol{\Delta}_{g}) = (4, 3, 1)'$ and 
$$\boldsymbol{\Gamma}_g=\begin{pmatrix}
0.36 & 0.48 & -0.80 \\ 
-0.80 & 0.60 & 0.0 \\ 
0.48 & 0.64 & 0.6 \\ 
 \end{pmatrix},$$
for $g=1,2,3$.
The simulated components were well separated to show parameter recovery. MSPE was run on 100 such datasets and perfect classification obtained on this well separated data each time. Note that there is an identifiability issue with individual parameter estimates (different combinations of individual parameter estimates yield the same fit) and closed form equations for overall mean and variance are not available. Hence, we demonstrate parameter recovery of overall cluster-specific mean and covariances in Table~\ref{tab:supp:parameterrecovery} by comparing estimates from data simulated (via a Metropolis-Hastings rule) using individual parameter estimates from the GEM fit. Clearly, the estimates are overall close to the generated values.}

\begin{table}[ht]
\centering
\caption{Parameter recovery of cluster-specific means and covariance matrices (rounded to two decimals)}
\resizebox{\textwidth}{!}{%
\begin{tabular}{lccc}
\hline
Parameter & Average generated & Average estimated & Mean squared error\\
\hline
$\text{Mean}_1$ & $(11.46, 12.14, -0.06)'$ & $(11.45, 12.08, -0.03)'$ & $(0, 0.02, 0.01)'$\\
$\text{Mean}_2$ & $(-1.96, -8.99, 0.27)'$ & $(-1.96, -8.97, 0.22)'$ & $(0, 0.01, 0.01)'$\\
$\text{Mean}_3$ & $(3.50, 1.47, 0.14)'$ & $(3.55, 1.51, 0.14)'$ & $(0, 0.00, 0.00)'$\\
$\text{Covariance}_1$ & $\begin{bmatrix}
3.35 & -1.67 & 2.21\\
-1.67 & 3.69 & -0.98\\
2.21 & -0.98 & 4.87\\
\end{bmatrix}$
& $\begin{bmatrix}
3.90 & -2.41 & 2.76\\
-2.41 & 5.00 & -1.59\\
2.76 & -1.59 & 6.90\\
\end{bmatrix}$
& $\begin{bmatrix}
0.47 & 0.75 & 0.47\\
0.75 & 2.10 & 0.52\\
0.47 & 0.52 & 4.62\\
\end{bmatrix}$\\

$\text{Covariance}_2$ & $\begin{bmatrix}
1.86 & -1.52 & 1.54\\
-1.52 & 4.92 & -0.90\\
1.54 & -0.90 & 3.99\\
\end{bmatrix}$
& $\begin{bmatrix}
2.31 & -2.22 & 2.06\\
-2.22 & 6.40 & -1.35\\
2.06 & -1.35 & 5.46\\
\end{bmatrix}$
& $\begin{bmatrix}
0.25 & 0.62 & 0.35\\
0.62 & 2.62 & 0.33\\
0.35 & 0.33 & 2.38\\
\end{bmatrix}$\\

$\text{Covariance}_3$ & $\begin{bmatrix}
0.42 & -0.36 & 0.35\\
-0.36 & 1.11 & -0.21\\
0.35 & -0.21 & 0.91\\
\end{bmatrix}$
& $\begin{bmatrix}
0.48 & -0.36 & 0.38\\
-0.36 & 1.19 & -0.22\\
0.38 & -0.22 & 0.97\\
\end{bmatrix}$
& $\begin{bmatrix}
0 & 0.00 & 0.00\\
0 & 0.01 & 0.00\\
0 & 0.00 & 0.01\\
\end{bmatrix}$\\

\hline
\end{tabular}}
\label{tab:supp:parameterrecovery}
\end{table}

\subsection{Performance on Additional Datasets}
In addition to those considered in the main body of the manuscript, we also considered the following data available through various {\sf R} packages:
\paragraph{Wine Dataset:} The expanded twenty seven variable {\tt wine} dataset, obtained from \texttt{pgmm} \citet{pgmm}, has 27 different measurements of chemical aspects of 178 Italian wines (three types of wine). 
\paragraph{Iris Dataset:} The {\tt iris} dataset (included with {\sf R}) consists of 150 observations, 50 each of 3 different species of iris. There are four different variables that were measured, namely the petal length and width and the sepal length and width.
\paragraph{Swiss Banknote Dataset:} The Swiss {\tt banknote} dataset, obtained from \texttt{MixGHD} \citep{mixGHDfromR} looked at 6 different measurements from 100 genuine and 100 counterfeit banknotes. The measurements were length, length of the diagonal, width of the right and left edges, and the top and bottom margin widths.
\paragraph{Crabs Dataset:} The {\tt crabs} dataset, obtained from \texttt{MASS} \citep{venables2002} contains 200 observations with 5 different variables that measure characteristics of crabs. There were 100 males and 100 females, and two different species of crabs, orange and blue. This creates four different groups of crabs based on gender/species combinations.
\paragraph{Bankruptcy Dataset:} The {\tt bankruptcy} dataset, obtained from \texttt{MixGHD} \citep{mixGHDfromR} looked at the ratio of retained earnings to total assets, and the ratio of earnings before interests and taxes to total assets of 33 financially sound and 33 bankrupt American firms. 
\paragraph{Yeast Dataset:} A subset of the {\tt yeast} dataset from \cite{nakai91,nakai92} sourced through the {\tt MixSAL} package \citep{MixSAL} is also used. There are measurements on three variables: McGeoch’s method for signal sequence recognition, the score of the ALOM membrane spanning region prediction program, and the score of discriminant analysis of the amino acid content of vacuolar and extracellular proteins along with the possible two cellular localization sites, CYT (cytosolic or cytoskeletal) and ME3 (membrane protein, no N-terminal signal) for the proteins. 
\paragraph{Diabetes Dataset:} The {\tt diabetes} dataset, obtained from \texttt{mclust}  \citep{fraley2012}  considered 145 non-obese adult patients with different types of diabetes classified as normal, overt and chemical. There were three measurements, the area under the plasma glucose curve, the area under the plasma insulin curve and the steady state plasma glucose.

\paragraph{Unsupervised classification}

Unsupervised classification, i.e., clustering, is performed on the (scaled) datasets mentioned above using the same comparison distributions as on the simulated data. 
The ARI and the number of groups chosen by the BIC is shown in Table~\ref{tab:supp:unsup}.

\begin{table}[ht]
	\centering
	\caption{Performance comparison of four mixture model families on real data for the unsupervised scenarios. Sample size, dimensionality, and the number of known groups (i.e., classes) are in parenthesis following each dataset name. For each implementation, the ARI and the number of selected components as well as the BIC are provided in parenthesis.}
	\label{tab:supp:unsup}
\resizebox{\textwidth}{!}{%
\begin{tabular}{p{6.4cm}llll}
\hline
Data & MSPE & MPE & {\tt ghpcm} & {\tt gpcm}\\
\hline
Banknote ($n=200, p=6, G=2$) & 0.86 (3; $-2681$) & 0.68 (4; $-2651$) & \textbf{0.98} (2; $-2700$) & 0.68 (4; $-2652$)\\
Bankruptcy ($n=66, p=2, G=2$) & 0.56 (4; $-235$) & 0.53 (3; $-228$) & 0 (3; $-236$) & \textbf{0.58} (3; $-263$)\\

Crabs ($n=200, p=5, G=4$) & 0.67 (3; 72) & \textbf{0.86} (4; 92) & 0.69 (3; 93) & 0.61 (3; 63)\\
Diabetes ($n=145, p=3, G=3$) & 0.44 (2; $-479$) & \textbf{0.66} (3; $-487$) & 0.4 (2; $-465$) & \textbf{0.66} (3; $-483$)\\

Iris ($n=150, p=4, G=3$) & 0.57 (2; $-799$) & \textbf{0.92} (3; $-795$) & 0.7 (4; $-582$) & 0.57 (2; $-791$)\\

Wine ($n=178, p=27, G=3$) & 0.83 (3; $-12179$) & 0.83 (3; $-11982$) & 0.81 (4; $-8665$) & \textbf{1} (3; $-11897$)\\
Yeast ($n=626, p=3, G=2$) & \textbf{0.49} (3; $-5042$) & 0.4 (4; $-5028$) & $-$0.04 (2; $-4863$) & 0.4 (4; $-5053$)\\
\hline
\end{tabular}
}
\label{tab:supp:unsup}
\end{table}

\edit{The {\tt banknote} data is interesting in that while the two elliptical mixtures, MPE and Gaussian mixture models, split the counterfeit and genuine banknotes into four different groups with the same classification overall (Table~\ref{tab:supp:classtab}), the selected MSPE model fits three components splitting the counterfeit banknotes into a larger and a smaller component, while the {\tt ghpcm} model splits the observations into two groups only. 
We see that, for the {\tt crabs} dataset, the MPE distribution exhibits the best performance (with eight ``blue males'' being misclassified into a different component) while the other three methods choose three components; however, the clusters found are a little different (Table~\ref{tab:supp:classtab}). For example, the MSPE model perfectly separates one species of crab from the other; however, for the ``blue'' species, it does not differentiate between the sexes. For the second species, there are only four misclassifications for differentiating the sexes. The {\tt ghpcm} model has a similar fit to the MSPE model (species separated nicely but not sexes), but with two fewer misclassifications. The {\tt gpcm} had a poorer fit compared to known labels, clustering sex better than species.}
\edit{The {\tt bankruptcy} data shows some interesting results. The {\tt ghpcm} model fits three clusters to the data, with two small clusters of two and three observations, respectively, with poor performance compared to known labels (Table~\ref{tab:supp:classtab}). The other three comparators performed somewhat similarly with MSPE fitting four components (including one with eight tightly clustered points). For the {\tt diabetes} data, the two best fitting skewed mixtures under-fit the number of components (seem to combine the normal and chemical classes but able to differentiate from the overt class) as compared to the elliptical mixtures, which fare better and similarly. Moreover, the estimates of skewness were not trivial, and both groups had heavier tails in the MSPE fit. On the other hand, for the MPE fit, the tails were approximately Gaussian (common $\beta_{g} \approx 1$). 
For the {\tt yeast} dataset, apart from the {\tt ghpcm} mixtures, the other three mixtures overfit the number of components; however, the {\tt ghpcm} mixture clustering was not meaningful compared to known labels.
For the {\tt iris} data, only the MPE mixtures' selected model had three components. While for the expanded 27-dimensional {\tt wine} dataset, the {\tt gpcm} mixtures perform best with perfect classification. Interestingly, the {\tt gpcm} mixtures have poorer performance relatively in the semi-supervised fits on these data. Note that this phenomenon, whereby cluster analysis can obtain better results compared to using semi-supervised classification, has been noted before, e.g., by \cite{vrbik15} and \cite{gallaugher17}. 
}
The relative performance of the MPE versus MSPE mixtures in Table~\ref{tab:supp:semisup} suggests that there are cases in which using these skewed mixtures might not be ideal, and could be a possible subject of future work. %
\begin{table}[ht]
\centering
\caption{Classification tables for the cluster analysis of each model and dataset.}
\scalebox{0.74}{
\begin{tabular}{ccccc}
\hline
Dataset& MSPE & MPE & {\tt ghpcm} & {\tt gpcm}\\
\hline
\\
{\tt banknote}&\begin{tabular}{rrrr}
\hline
&1 & 2 & 3\\
\hline
counterfeit&0 & 85 & 15\\
genuine&99 & 0 & 1\\
\hline
\end{tabular}&\begin{tabular}{rrrr}
\hline
1 & 2 & 3 & 4\\
\hline
85 & 0 & 0 & 15\\
0 & 75 & 24 & 1\\
\hline
\end{tabular}
&\begin{tabular}{rr}
\hline
1 & 2\\
\hline
100 & 0\\
1 & 99\\
\hline
\end{tabular}&\begin{tabular}{rrrr}
\hline
1 & 2 & 3 & 4\\
\hline
85 & 0 & 0 & 15\\
0 & 75 & 24 & 1\\
\hline
\end{tabular}\\ \\

{\tt bankruptcy} &\begin{tabular}{rrrrr}
\hline
&1 & 2 & 3 & 4\\
\hline
bankruptcy&30 & 0 & 3 & 0\\
financially sound&0 & 15 & 10 & 8\\
\hline
\end{tabular}
&\begin{tabular}{rrr}
\hline
1 & 2 & 3\\
\hline
0 & 21 & 12\\
28 & 5 & 0\\
\hline
\end{tabular}&\begin{tabular}{rrr}
\hline
1 & 2 & 3\\
\hline
31 & 0 & 2\\
30 & 3 & 0\\
\hline
\end{tabular}&\begin{tabular}{rrr}
\hline
1 & 2 & 3\\
\hline
0 & 22 & 11\\
29 & 4 & 0\\
\hline
\end{tabular}\\ \\

{\tt crabs} &\begin{tabular}{lrrr}
\hline
&1 & 2 & 3\\
\hline
Blue males&50 & 0 & 0\\
Blue females&50 & 0 & 0\\
Orange males&0 & 50 & 0\\
Orange females&0 & 4 & 46\\
\hline
\end{tabular}
&\begin{tabular}{rrrr}
\hline
1 & 2 & 3 & 4\\
\hline
8 & 0 & 0 & 42\\
49 & 0 & 1 & 0\\
0 & 50 & 0 & 0\\
0 & 2 & 48 & 0\\
\hline
\end{tabular}&\begin{tabular}{rrr}
\hline
1 & 2 & 3\\
\hline
50 & 0 & 0\\
50 & 0 & 0\\
0 & 50 & 0\\
0 & 2 & 48\\
\hline
\end{tabular}&\begin{tabular}{rrr}
\hline
1 & 2 & 3\\
\hline
38 & 12 & 0\\
0 & 49 & 1\\
50 & 0 & 0\\
2 & 0 & 48\\
\hline
\end{tabular}
\\ \\

{\tt diabetes}&\begin{tabular}{lrr}
\hline
&1 & 2\\
\hline
Chemical&34 & 2\\
Normal&76 & 0\\
Overt&3 & 30\\
\hline
\end{tabular}&\begin{tabular}{rrr}
\hline
1 & 2 & 3\\
\hline
9 & 26 & 1\\
72 & 4 & 0\\
0 & 6 & 27\\
\hline
\end{tabular}&\begin{tabular}{rr}
\hline
1 & 2\\
\hline
34 & 2\\
76 & 0\\
5 & 28\\
\hline
\end{tabular}&\begin{tabular}{rrr}
\hline
1 & 2 &3\\
\hline
9 & 26 & 1\\
72 & 4 & 0\\
0 & 6 & 27\\
\hline
\end{tabular}\\
\\

{\tt iris}&\begin{tabular}{lrr}
\hline
&1 & 2\\
\hline
{\it setosa}&0 & 50\\
{\it versicolor}&50 & 0\\
{\it virginica}&50 & 0\\
\hline
\end{tabular}&\begin{tabular}{rrr}
\hline
1 & 2 & 3\\
\hline
0 & 50 & 0\\
4 & 0 & 46\\
50 & 0 & 0\\
\hline
\end{tabular}&\begin{tabular}{rrrr}
\hline
1 &2& 3 & 4\\
\hline
0 & 0 & 29 & 21\\
46 & 4 & 0 & 0\\
5 & 45 & 0 & 0\\
\hline
\end{tabular}&\begin{tabular}{rr}
\hline
1 & 2\\
\hline
0 & 50\\
50 & 0\\
50 & 0\\
\hline
\end{tabular}\\ \\

{\tt wine}&\begin{tabular}{lrrr}
\hline
&1&2 & 3\\
\hline
Barolo&58 & 1 & 0\\
Grignolino&7 & 62 & 2\\
Barbera&0 & 0 & 48\\
\hline
\end{tabular}&\begin{tabular}{rrr}
\hline
1 & 2 & 3\\
\hline
58 & 1 & 0\\
7 & 62 & 2\\
0 & 0 & 48\\
\hline
\end{tabular}&\begin{tabular}{rrrr}
\hline
1 & 2 & 3 & 4\\
\hline
0 & 59 & 0 & 0\\
63 & 1 & 0 & 7\\
0 & 0 & 29 & 19\\
\hline
\end{tabular}
&\begin{tabular}{rrr}
\hline
1 & 2 & 3\\
\hline
0 & 59 & 0\\
71 & 0 & 0\\
0 & 0 & 48\\
\hline
\end{tabular}\\ \\

{\tt yeast}&\begin{tabular}{lrrr}
\hline
&1 & 2 & 3\\
\hline
CYT&354 & 33 & 76\\
ME3&10 & 152 & 1\\
\hline
\end{tabular}&\begin{tabular}{rrrr}
\hline
1 & 2 & 3 & 4\\
\hline
319 & 3 & 100 & 41\\
23 & 135 & 0 & 5\\
\hline
\end{tabular}&\begin{tabular}{rr}
\hline
1 & 2\\
\hline
291 & 172\\
155 & 8\\
\hline
\end{tabular}&\begin{tabular}{rrrr}
\hline
1 & 2 & 3 & 4\\
\hline
307 & 5 & 128 & 23\\
20 & 135 & 0 & 8\\
\hline
\end{tabular}
\\ \\

\hline
\end{tabular}}
\label{tab:supp:classtab}
\end{table}

\paragraph{Semi-supervised classification} For each dataset, we take 25 labelled/unlabelled splits with 25\% supervision. In Table \ref{tab:supp:semisup}, we display the median ARI values along with the first and third quartiles over the 25 splits. 
For the most part, as found in the main text of this manuscript, performance in the semi-supervised scenarios was better than in the fully unsupervised scenarios. \edit{Performance across the four comparators was also quite comparable with few exceptions. }
\edit{For the {\tt yeast} data, both in the unsupervised and semi-supervised context, the MSPE mixtures performed the best. Similarly, for the {\tt diabetes} data, the MSPE mixtures perform the best, with {\tt gpcm} mixtures having the most inferior classification performance.  
On the 27 dimensional {\tt wine} data, the MPE mixtures performed well with MSPE mixtures having more variability in ARI across the runs. 
For the {\tt iris} dataset, the MPE and {\tt ghpcm} models showed the best overall performance while, for the {\tt banknote} dataset, all algorithms exhibit similar performance. 
For the {bankruptcy} data, the {\tt gpcm} algorithm performed the poorest while, for {\tt crabs}, the {\tt gpcm} algorithm performed the best along with {\tt ghpcm} and MPE mixtures which had similar performance. 
For the {\tt crabs} dataset, although the MSPE models had poorer classification compared to the other three mixtures, the performance was still close to the other mixture distributions.  
}
A reviewer noted that using mixtures of canonical fundamental skew $t$ (CFUST) distributions \citep{lee2016}, one can obtain an ARI close to 1 with there being only one misclassification. However, we were unable to obtain this solution from CFUST; perhaps due to different initialization. \edit{For the semi-supervised runs, this solution could be obtained for all four comparator distributions (for the best out of 25 runs). }

\begin{table}[ht]
	\centering
	\caption{Median ARI values along with first and third quartiles in parenthesis for the four different models for each dataset for the semi-supervised runs. Size, dimensionality and the number of known groups (i.e., classes) are in parenthesis following each dataset name.}
	\label{tab:supp:semisup}
	\resizebox{\textwidth}{!}{%
\begin{tabular}{p{6.4cm}llll}
\hline
Data & MSPE & MPE & ghpcm & gpcm\\
\hline
Banknote ($n=200, p=6, G=2$) & 0.97 (0.97, 0.97) & \textbf{0.97 (0.97, 1)} & 0.97 (0.97, 0.97) & 0.97 (0.97, 0.97)\\

Bankruptcy ($n=66, p=2, G=2$) & 0.77 (0.63, 0.84) & \textbf{0.77 (0.7, 0.84)} & 0.77 (0.7, 0.77) & 0.51 (0.4, 0.77)\\

Crabs ($n=200, p=5, G=4$) & 0.8 (0.78, 0.83) & 0.85 (0.83, 0.87) & 0.85 (0.82, 0.86) & \textbf{0.86 (0.83, 0.88)}\\
Diabetes ($n=145, p=3, G=3$) & \textbf{0.74 (0.7, 0.79)} & 0.73 (0.68, 0.79) & 0.73 (0.7, 0.79) & 0.68 (0.67, 0.73)\\

Iris ($n=150, p=4, G=3$) & 0.9 (0.89, 0.92) & 0.92 (0.9, 0.93) & \textbf{0.92 (0.9, 0.95)} & 0.9 (0.87, 0.93)\\
Wine ($n=178, p=27, G=3$) & 0.91 (0.87, 0.95) & \textbf{0.95 (0.95, 0.96)} & 0.93 (0, 0.95) & 0.49 (0.44, 0.62)\\

Yeast ($n=626, p=3, G=2$) & \textbf{0.84 (0.83, 0.86)} & 0.78 (0.75, 0.81) & 0.74 (0.71, 0.76) & 0.81 (0.78, 0.83)\\
\hline
\end{tabular}	} 
\label{tab:supp:semisup}
\end{table}

\end{document}